\documentstyle[11pt]{article}

\makeatletter
\@addtoreset{equation}{section}
\makeatother

\def\dalemb#1#2{{\vbox{\hrule height .#2pt
        \hbox{\vrule width.#2pt height#1pt \kern#1pt
                \vrule width.#2pt}
        \hrule height.#2pt}}}

    \let\e=\epsilon
  \let\q=\theta  
  \let\n=\nu

\let\la=\label  
   
\def\nn{\nonumber} \def\bd{\begin{document}} \def\ed{\end{document}}
\def\ds{\documentstyle} \let\fr=\frac \let\bl=\bigl \let\br=\bigr
\let\Br=\Bigr \let\Bl=\Bigl 
\let\bm=\bibitem
\let\na=\nabla
\let\pa=\partial \let\ov=\overline
\def\ie{{\it i.e.\ }} 
\newcommand{\be}{\begin{equation}} 
\newcommand{\ee}{\end{equation}} 
\def\ba{\begin{array}}
\def\ea{\end{array}}
\def\ft#1#2{{\textstyle{{\scriptstyle #1}\over {\scriptstyle #2}}}}
\def\fft#1#2{{#1 \over #2}}
\def\del{\partial}
\def\sst#1{{\scriptscriptstyle #1}}
\def\oneone{\rlap 1\mkern4mu{\rm l}}
\def\e7{E_{7(+7)}}
\def\td{\tilde}
\def\wtd{\widetilde}
\def\im{{\rm i}}
\def\bog{Bogomol'nyi\ }
\def\q{{\tilde q}}
\def\hast{{\hat\ast}}
\def\0{{\sst{(0)}}}
\def\1{{\sst{(1)}}}
\def\2{{\sst{(2)}}}
\def\3{{\sst{(3)}}}
\def\4{{\sst{(4)}}}
\def\5{{\sst{(5)}}}
\def\6{{\sst{(6)}}}
\def\7{{\sst{(7)}}}
\def\8{{\sst{(8)}}}
\def\n{{\sst{(n)}}}

\def\hA{\hat{\cal A}}
\def\ns{{\sst {\rm NS}}}
\def\rr{{\sst {\rm RR}}}
\def\tH{{\widetilde H}}
\def\tB{{\widetilde B}}
\def\cA{{\cal A}}
\def\cF{{\cal F}}
\def\tF{{\wtd F}}
\def\v{{{\cal V}}}
\def\Z{\rlap{\sf Z}\mkern3mu{\sf Z}}
\def\ep{{\epsilon}}
\def\IIA{{\rm IIA}}
\def\IIB{{\rm IIB}}
\def\ads{{\rm AdS}}
\def\R{\rlap{\rm I}\mkern3mu{\rm R}}

\newcommand{\ho}[1]{$\, ^{#1}$}
\newcommand{\hoch}[1]{$\, ^{#1}$}
\newcommand{\bea}{\begin{eqnarray}} 
\newcommand{\eea}{\end{eqnarray}} 
\newcommand{\ra}{\rightarrow}
\newcommand{\lra}{\longrightarrow}
\newcommand{\Lra}{\Leftrightarrow}
\newcommand{\ap}{\alpha^\prime}
\newcommand{\bp}{\tilde \beta^\prime}
\newcommand{\tr}{{\rm tr} }
\newcommand{\Tr}{{\rm Tr} } 
\newcommand{\NP}{Nucl. Phys. }
\newcommand{\tamphys}{\it Center for Theoretical Physics,
Texas A\&M University, College Station, Texas 77843}

\newcommand{\auth}{M.J. Duff}

\begin{document}
\begin{flushright}
\hfill{CTP-TAMU-30/98}\\
\hfill{hep-th/9808100}\\
\end{flushright}
\vspace{24pt}

\begin{center}
{ \large 
{\bf Anti-de Sitter space, branes, singletons, superconformal field 
theories and all that\footnote{Based on talks delivered at the the 
{\it PASCOS 98} conference, Northeastern University, March 1998; the {\it 
Superfivebranes and Physics in $5+1$ dimensions} conference, ICTP, 
Trieste, Italy, April 1998; the {\it Arnowitt Fest}, Texas A\&M 
University, April 1998; the {\it Strings 98} conference, ITP, 
Santa Barbara, June 1998.  Research supported in part by NSF 
Grant PHY-9722090.}}.}

\vspace{36pt}

\auth

\vspace{10pt}

{\tamphys}

\vspace{44pt}

\underline{ABSTRACT}

\end{center}
There has recently been a revival of interest in anti de-Sitter space 
(AdS) brought about by the conjectured duality beteeen physics in the 
bulk of AdS and a conformal field theory on the boundary. Since the whole 
subject of branes, singletons and superconformal field theories 
on the AdS boundary was an active area of research about ten years ago, I begin 
with a historical review, including the ``Membrane at the end of the 
universe'' idea.
Next I discuss two recent papers with Lu and Pope  
on $AdS_{5}¥ \times S^{5}¥$ and on 
$AdS_{3}¥ \times S^{3}¥$, respectively. In each case we note 
that  odd-dimensional spheres $S^{{2n+1}}¥$ may be regarded as  $U(1)$ 
bundles over $CP^{n}¥$ and that this permits an unconventional ``Hopf''
duality along the $U(1)$ fibre.  This leads in particular to the phenomenon   
of {\it BPS without BPS} whereby states which appear to be non-BPS in 
one picture are seen to be BPS in the dual picture.

{\vfill\leftline{}\vfill
\vskip	10pt

\pagebreak
\setcounter{page}{1}

\section{Historical review}

\subsection{Gauged extended supergravities and their Kaluza-Klein origin}

In the early 80's there was great interest in 
four-dimensional $N$-extended supergravities for which the global $SO(N)$ is 
promoted to a gauge
symmetry \cite{Das}. In these theories the underlying supersymmetry algebra is 
no longer 
Poincare but rather anti-de Sitter (AdS$_4$) and the
Lagrangian has a non-vanishing cosmological constant $\Lambda$ proportional to
the square of the gauge coupling constant $e$:  
\be
G\Lambda=-e^{2}¥
\la{G}
\ee
where $G$ is Newton's constant. The $N>4$ gauged supergravities were 
particularly interesting since the cosmological constant $\Lambda$ does 
not get renormalized \cite{CDGR} and hence the $SO(N)$ gauge symmetry has 
vanishing $\beta$-function\footnote{For $N\leq 4$, the 
$beta$ function (which receives a contribution from the spin $3/2$ 
gravitinos) is positive and the pure supergravity theories are not 
asymptotically free. The addition of matter supermultiplets only makes 
the $\beta$ function more positive \cite{Supergravity81} and hence 
gravitinos 
can
never be  confined. I am grateful to Karim Benakli, Rene Martinez Acosta 
and
Parid Hoxha  for discussions on this point.}.  The relation (\ref{G}) 
suggested
that there might be a Kaluza-Klein interpretation since in such theories 
the 
coupling constant of the gauge group arising from the isometries of 
the extra dimensions is given by
\be
e^{2}¥ \sim Gm^{2}¥
\la{e}
\ee
where $m^{-1}¥$ is the size of the compact space. Moreover, there is 
typically a negative cosmological constant
\be
\Lambda\sim -m^{2}¥
\la{Lambda}
\ee
Combining (\ref{e}) and (\ref{Lambda}), we recover (\ref{G}).  Indeed, 
the
maximal $(D=4,N=8)$ gauged supergravity \cite{DN} was seen to correspond 
to the 
massless sector of $(D=11,N=1)$
supergravity \cite{Cremmer} compactified on an $S^7$ whose metric admits 
an $SO(8)$
isometry and $8$ Killing spinors \cite{DP}. An important ingredient in 
these 
developments that
had been insufficiently emphasized in earlier work on Kaluza-Klein
theory was that the AdS$_4 \times S^7$ geometry was not fed in by hand
but resulted from a {\it spontaneous compactification}, i.e. the
vacuum state was obtained by finding a stable solution of the
higher-dimensional field equations \cite{CS}.  The mechanism of
spontaneous compactification appropriate to the AdS$_4 \times S^7$
solution of eleven-dimensional supergravity was provided by the 
Freund-Rubin mechanism \cite{FR} in which the $4$-form field strength in
spacetime $F_{\mu\nu\rho\sigma}$ ($\mu=0,1,2,3$) is proportional to
the alternating symbol $\epsilon_{\mu\nu\rho\sigma}$ \cite{DV}:
\be
F_{\mu\nu\rho\sigma}¥ \sim \epsilon_{\mu\nu\rho\sigma}
\la{F_{4}¥}
\ee
A summary of this $S^{7}¥$ and other $X^{7}¥$ compactifications of $D=11$ 
supergravity  down to $AdS_{4}$ may be found in
\cite{DNP}. By applying a similar mechanism to the $7$-form dual of this 
field
strength one could also find compactifications on AdS$_{7} \times
S^{4}$ \cite{PTV} whose massless sector describes gauged maximal
$N=4$, $SO(5)$ supergravity in $D=7$ \cite{PPV,TV}. Type IIB supergravity 
in $D=10$, with its self-dual
$5$-form field strength, also admits a Freund-Rubin compactification
on AdS$_{5}\times S^{5}$ \cite{Schwarz,GM,KRV} whose massless sector 
describes
gauged maximal $N=8$ supergravity in $D=5$ \cite{PPV2,GRW}.

\bigskip\bigskip
\begin{center}
\begin{tabular}{ccc}
{\bf Compactification}&{\bf Supergroup}&{\bf Bosonic~subgroup}\\
$AdS_{4}¥\times S^{7}$¥&$OSp(4|8)$&$SO(3,2) \times SO(8)$\\
$AdS_{5}¥\times S^{5}$¥&$SU(2,2|4)$&$SO(4,2) \times SO(6)$\\
$AdS_{7}¥\times S^{4}$¥&$OSp(6,2|4)$&$SO(6,2) \times SO(5)$
\end{tabular}
\label{supergroups}
\end{center}
\medskip
\centerline{Table 1: Compactifications and their symmetries.}
In the three cases given above, the symmetry of the vacuum is
described by the supergroups $OSp(4|8)$, $SU(2,2|4)$ and $OSp(6,2|4)$
for the $S^7$, $S^5$ and $S^4$ compactifications respectively, as 
shown in Table 1.

\subsection{Singletons}
\la{singletons}

Each of these groups is known to admit the so-called singleton, doubleton 
or 
tripleton\footnote{Our nomenclature is based on the 
$AdS_{4}¥,AdS_{5}¥$ and $AdS_{7}¥$ groups having ranks $2,3$ and $4$, 
respectively,  and differs from that of Gunaydin.} 
supermultiplets \cite{Gunaydin1} as shown in Table 2. 
\bigskip\bigskip
\begin{center}
\begin{tabular}{ccc}
{\bf Supergroup}&{\bf Supermultiplet}&{\bf Field~content}\\
$OSp(4|8)$&$(n=8,d=3)$~singleton&8~scalars,8~spinors\\
$SU(2,2|4)$&$(n=4,d=4)$~doubleton&1~vector,4~spinors,6~scalars\\
$OSp(6,2|4)$&$((n_{+}¥,n_{-}¥)=(2,0),d=6)$
~tripleton&1~chiral~2-form,8~spinors,5~scalars
\end{tabular}
\label{fields}
\end{center}
\medskip
\centerline{Table 2: Superconformal groups and their singleton, doubleton 
and 
tripleton repesentations.}
\bigskip\bigskip
We recall that 
singletons are those strange representations of $AdS$ first identified 
by Dirac \cite{Dirac} which admit no analogue in flat spacetime. They 
have been much studied by Fronsdal and collaborators 
\cite{Fronsdal,Flato}.  Let us first consider $AdS_{4}$ which can be 
defined as the 
four-dimensional hyperboloid 
\be
\eta_{ab}y^{a}y^{b}=-\frac{1}{a^{2}}
\ee
in $R^{5}$ with Cartesian coordinates $y^{a}¥$ where
\be
\eta_{ab}¥ =diag(-1,1,1,1,-1)
\ee
In polar coordinates $x^{\mu}¥=(t,r,\theta,\phi)$ the line element 
may be written
\be
g_{\mu\nu}dx^{\mu}dx^{\nu}
=-(1+a^{2}¥r^{2}¥)dt^{2}¥+(1+a^{2}¥r^{2}¥)^{-1}¥dr^{2}¥
+r^{2}¥(d\theta^{2}¥+sin^{2}¥ \theta d\phi^{2}¥)
\la{metric}
\ee
Representations of $SO(3,2)$ are denoted $D(E_{0},s)$, where $E_{0}¥$ 
is the lowest energy eigenvalue (in units of $a$) and $s$ is the 
total angular momentum. The representation is unitary provided 
$E_{0}¥\geq s+1/2$ for $s=0,1/2$ and $E_{0}¥\geq s+1$ for $s \geq 1$. 
The representations are all infinite dimensional. In the 
supersymmetric context, all linear irreducible representations of 
$N=1$ $AdS$ supersymmetry were classified by Heidenreich 
\cite{Heidenreich}. They fall into 4 classes:

1. $D(1/2,0) \oplus D(1,1/2)$ 
 
2. $D(E_{0}¥,0) \oplus D(E_{0}¥+1/2,1/2) \oplus D(E_{0}¥+1,0), E_{0}¥
\geq 1/2$

3. $D(s+1,s) \oplus D(s+3/2,s+1/2), s\geq 1/2$

4. $D(E_{0}¥,s) \oplus D(E_{0}¥+1/2,s+1/2) \oplus D(E_{0}¥+1/2,s-1/2) 
\oplus D(E_{0}¥+1,s).$

Class 1 is the singleton supermultiplet which has no analogue in Poincare
supersymmetry. Class 2 is the Wess-Zumino supermultiplet. Class 3 is the 
gauge
supermultiplet with spins $s$ and $s+1/2$ with $s\geq1/2$. Class 4 is the
higher spin supermultiplet. The corresponding study of $OSp(4|N)$
representations was neglected in the literature until their importance in
Kaluza-Klein supergravity became apparent. For example, the
round $S^7$ leads to massive $N=8$ supermultiplets with maximum spin $2$. 
This corresponds 
to 
an
$AdS$ type of multiplet shortening analogous to the shorteneing due to 
central
charges in Poincare supersymmetry \cite{FN}. Two features emerge: (1)  
$OSp(4|N)$ multiplets may be decomposed into $OSp(4|1)$ mutiplets discussed
above; (2) In the limit as $a \rightarrow 0$ and the $OSp(4|N)$ contracts 
to 
the
$N$-extended Poincare algebra, all short $AdS$ multiplets become massless
Poincare multiplets.

\subsection{Singletons live on the boundary}

As emphasized by Fronsdal et al \cite{Fronsdal,Flato}, singletons are best 
thought of as 
living not in the $(d+1)$-dimensional bulk of the $AdS_{d+1}¥$ spacetime 
but rather on the $d$-dimensional $S^{1} \times S^{d-1}$ boundary where the 
$AdS$ group $SO(d-1,2)$ plays the role of the {\it conformal} group. 
Remaining 
for the moment 
with our $4$-dimensional example, consider a scalar field 
$\Phi(t,r,\theta,\phi)$ on $AdS_{4}¥$ with metric (\ref{metric}), 
described by 
the action
\be
S_{bulk}=-\frac{1}{2}\int_{AdS_{4}}d^{4}¥x \sqrt{-g} 
\Phi(-g^{\mu\nu}¥\nabla_{\mu}¥ \nabla_{\nu}¥+M^{2}¥)\Phi
\la{bulk}
\ee
Note that this differs from the conventional Klein-Gordon action by a 
boundary term. Since the scalar Laplacian on $AdS_{4}¥$ has eigenvalues 
$E_{0}(E_{0}-3)a^{2}$, the critical value of $M^{2}¥$ for a singleton with 
$(E_{0},s)=(1/2,0)$ is
\be
M^{2}=\frac{5}{4}a^{2}¥
\ee
In this case, one can show with some effort \cite{Fronsdal,Flato} that as 
$r 
\rightarrow \infty$,
\be
\Phi(t,r,\theta,\phi) \rightarrow 
r^{{-1/2}}\phi(t,\theta,\phi)
\ee
and hence that the radial dependence drops out:
\be
S_{boundary}¥=-\frac{1}{2}\int_{S^{1} \times S^{2}}d^{3}¥\xi 
\sqrt{-h}[h^{ij}\nabla_{i}\phi \nabla_{j}\phi +\frac{1}{4}a^{2}\phi^{2}]
\la{boundary}
\ee
Here we are integrating over a $3$-manifold with $S^{1} \times S^{2}$ 
topology and with metric
\be
h^{ij}¥d\xi^{i}¥d\xi^{j}¥=-dt^{2} + 
\frac{1}{a^{2}¥}(d\theta^{2}¥+sin^{2}¥ \theta d\phi^{2}¥)
\la{boundarymetric}
\ee
This $3$-manifold is sometimes referred to as the {\it boundary} of 
$AdS_{4}$ 
but note that the metric $h_{ij}$ is not obtained by taking the 
$r \rightarrow \infty$ limit of $g_{\mu\nu}$ but rather the $r \rightarrow 
\infty$ limit of the conformally rescaled metric $\Omega^{2}¥g_{\mu\nu}$ 
where $\Omega=1/ar$. The radius of the $S^{2}¥$ is $a^{-1}¥$ not 
infinity. Most particle physicists are familiar with the conformal 
group in flat Minkowski space. It is the group of coordinate 
transformations which leave invariant the Minkowski lightcone. In the 
case of three-dimensional Minkowski space, $M_{3}$, it is $SO(3,2)$. 
In the present context, however, the spacetime is curved with 
topology $S^{1} \times S^{2}¥$, but still admits $SO(3,2)$ as its 
conformal group\footnote{One sometimes finds the statement in 
the physics literature that the only compact spaces admitting 
conformal Killing vectors are those isomorphic to spheres. By a 
theorem of Yano and Nagano \cite{Yano}, this is true for {\it 
Einstein spaces}, but $S^{1} \times S^{2}¥$ is not Einstein.}, i.e. 
as the group which leaves invariant the 
three-dimensional lightcone $h_{ij}¥ 
d\xi^{i}¥d\xi^{j}¥=0$.  The 
failure to discriminate between these different kinds of conformal 
invariance is, we believe, a source of confusion in the singleton 
literature. In particular, the $\phi^{2}¥$ ``mass'' term appearing in 
the action (\ref{boundary}) would be incompatible with conformal 
invariance if the action were on $M_{3}¥$ but is essential for 
conformal invariance on $S^{1} \times S^{2}$. Moreover, the 
coefficient $a^{2}¥/4$ is uniquely fixed \cite{BD}.

So although singleton actions of the form (\ref{bulk}) and their 
superpartners appeared in the Kaluza-Klein harmonic expansions on $AdS_{4} 
\times S^7$ \cite{Sezgin,NS,GRW2}, they could be
gauged away everywhere except on the boundary where the above 
$OSp(4|8)$ 
corresponds to the superconformal group \cite{Nahm}. One 
finds an $(n=8,d=3)$ supermultiplet with $8$ scalars $\phi^{A}$ and $8$ 
spinors
$\chi^{\dot{A}}$, where the indices $A$ and $\dot{A}$ range over $1$ to 
$8$ and denote the 
$8_{s}$ and $8_{c}$ representations of $SO(8)$, respectively. The 
$OSp(4|8)$ 
action is a generalization of (\ref{boundary}) and is given by \cite{BD} 
\be
S_{singleton}=-\frac{1}{2}\int_{S^{1} \times S^{2}}d^{3}¥\xi 
\sqrt{-h}[h^{ij}\nabla_{i}\phi^{A} \nabla_{j}\phi^{A} +\frac{1}{4}a^{2}
\phi^{A}\phi^{A}
+i{\bar{\chi}}^{\dot{A}}(1-\gamma))\gamma^{i}¥D_{i}¥\chi^{\dot{A}}]
\la{singletonaction}
\ee
where $\gamma=-\gamma_{0}¥\gamma_{1}¥\gamma_{2}¥$ and where $D_{i}$ is the 
covariant derivative appropriate to the $S^{1}¥ \times S^{2}¥$ background. 

In the
case of $AdS_{5} \times S^5$ one finds a $(n=4,d=4)$ supermultiplet with 
$1$ vector $A_{i}$, $(i=0,1,2,3)$, a complex spinor 
$\lambda^{a}{}_{+}$, $(a=1,2,3,4)$, obeying 
$\gamma_{5}\lambda^{a}{}_{+}=\lambda^{a}{}_{+}$ and $6$ real scalars 
$\phi^{ab}$, obeying $\phi^{ab}=-\phi^{ab}¥$, 
$\phi^{ab}=\epsilon^{abcd}¥ \phi_{cd}¥/2$.  The corresponding action for 
the 
doubletons of $SU(2,2|4)$ is 
\cite{NST}
\be  
S_{doubleton}= \int_{S^{1} \times S^{3}} 
[-\frac{1}{4}F_{ij}F^{ij} -\frac{1}{4}a^{2}¥\phi_{ab}\phi^{ab} 
-\frac{1}{4}\partial_{i}\phi_{ab}\partial^{i}\phi^{ab}
+i{\bar{\lambda}}_{+a}\gamma^{i}D_{i}\lambda^{a}{}_{+}]
\la{doubletonaction} 
\ee
where $F_{ij}=2\partial_{[i}¥A_{j]}¥$. However, in contrast to the 
singletons, 
we know of no derivation of this doubleton action on the boundary 
starting from an action in the bulk analogous to (\ref{boundary}).

In the case of $AdS_{7} \times S^4$ one finds a
$((n_+,n_-)=(2,0),d=6)$ supermultiplet with a $2$-form $B_{ij}$, 
$(i=0,1,\ldots5)$, whose field strength is self-dual, $8$ spinors 
$\lambda^{A}{}_{+}$, $(A=1,2,3,4)$, obeying 
$\gamma^{7}\lambda^{A}{}_{+}=\lambda^{A}{}_{+}$ and $5$ scalars 
$\phi^{a}$, $(a=1,2,\ldots 5)$. The $OSp(6,2|4)$ tripleton covariant field 
equations on $S^{1} \times S^{5}$ are \cite{NST}:
\[
(\nabla^{i}\nabla_{i}¥-4a^{2}¥)\phi^{a}¥=0
\]
\[
\gamma^{i}¥D_{i}¥\lambda^{A}{}_{+}=0
\]
\be
H_{ijk}=\frac{1}{3!}\sqrt{-h} \epsilon^{ijklmn}H_{lmn}
\la{tripletonequations}
\ee
where $H_{ijk}=3\partial_{[i}¥B_{jk]}¥$.  Once again, we know of no 
derivation 
of these tripleton field equations on the boundary starting from 
equations in 
the bulk.

\subsection{The membrane as a singleton: the membrane/supergravity 
bootstrap}
\la{bootstrap}

Being defined over the boundary of $AdS_{4}¥$, the $OSp(4|8)$ singleton 
action (\ref{singletonaction}) is a {\it three dimensional} theory 
with signature $(-,+,+)$ describing $8$ scalars and $8$ spinors.    
With the discovery of the eleven-dimensional supermembrane 
\cite{BST}, it was noted that $8$ scalars and $8$ spinors on a 
three-dimensional worldvolume with signature $(-,+,+)$ is just what 
is obtained after gauge-fixing the supermembrane action! Moreover, 
kappa-symmetry of this supermembrane action forces the  
background fields to obey the field equations of $(N=1,D=11)$ 
supergravity. It was therefore 
suggested \cite{Fifteen} that on the $AdS_{4} \times S^{7}¥$ 
supergravity background, the supermembrane could be regarded as the 
singleton of 
$OSp(4|8)$ whose worldvolume occupies the $S^{1}¥\times S^{2}¥$ boundary 
of 
the $AdS_{4}¥$.  Noting that these singletons also appear in the 
Kaluza-Klein harmonic expansion of this supergravity background, this 
further suggested a form of bootstrap \cite{Fifteen} in which the 
supergravity gives rise to the membrane on the boundary which in turn 
yields 
the 
supergravity in the bulk. This conjecture received further support with 
the subsequent discovery of the ``membrane at the end of the 
universe'' \cite{BDPS} to be discussed in section \ref{universe}, and 
the 
realisation \cite{dust} that the eleven-dimensional supermembrane 
emerges as a solution of the $D=11$ supergravity field equations. 

The possibility of a similar $3$-brane/supergravity bootstrap 
arising for the $SU(2,2|4)$ doubletons on $AdS_{5} \times 
S^{5}$ and a similar $5$-brane/supergravity bootstrap arising for the 
$OSp(6,2|4)$ tripletons on $AdS_{7} \times S^{4}¥$ was also 
considered \cite{Fifteen}.
Ironically, however, it was (erroneously as we now know) rejected 
since the only supermembranes that were known at the time \cite{AETW} had 
worldvolume theories described by {\it scalar} supermultiplets, 
whereas the doubletons and tripletons required vector and tensor 
supermultiplets, respectively. See section \ref{revisit}. 

Nevertheless, since everything seemed to fit nicely for the $(d=3,D=11)$ 
slot 
on the 
brane-scan of supersymmetric extended objects with worldvolume 
dimension $d$, there
followed a good deal of activity relating other super $p$-branes in 
other dimensions to singletons and superconformal field theories
\cite{BDPS,BDPS2,BD,Classical,BD2,BSTan,BSS,NST,DPS}. In 
particular, it was pointed out \cite{BD,NST,BD2} that there was a 
one-to one-correspondence between the $12$ points on the brane-scan 
as it was then known 
\cite{AETW} and the $12$ superconformal groups in Nahm's 
classification \cite{Nahm} admitting singleton 
representations, as shown in Table 3. The number of 
dimensions transverse to the brane, $D-d$, equals the number of scalars 
in the singleton supermultiplet. (The two factors appearing in the $d=2$ 
case
is simply a reflection of the ability of strings to have right and left 
movers.
For brevity, we have written the Type $II$ assignments in Table 3, but 
more
generally we could have $OSp(p|2) \times OSp(q|2)$ where $p$ and $q$ are 
the
number of left and right supersymmetries \cite{GNST}.) Note that the $d=6$ 
upper limit on the worldvolume dimension is consistent with the 
requirement of renormalizability \cite{Classical}.     

\begin{center}
\begin{tabular}{cccccccccc}
~&D$\uparrow$&&&&&&\\
~&11&.&~~~~~~~~~~~~~~~~~~&&${\bf OSp(8|4)}$&&&\\
~&10&.&~~~~~~~~&$[OSp(8|2)]^2$&&&&$OSp(6,2|2)$\\
~&~9&.&~~~~~~~~&&&&$F(4)$&\\
~&~8&.&~~~~~~~~&&&$SU(2,2|2)$&&\\
~&~7&.&~~~~~~~~&&$OSp(4|4)$&&&\\
~&~6&.&~~~~~~~~&$[OSp(4|2)]^2$&~&$SU(2,2|1)$&&\\
~&~5&.&~~~~~~~~&&$OSp(2|4)$&&&\\
~&~4&.&~~~~~~~~&$[OSp(2|2)]^2$&$OSp(1|4)$&&&\\
~&~3&.&~~~~~~~~&$[OSp(1|2)]^2$&~&&&\\
~&~2&.&~~~~~~~~&&&&&&\\
~&~1&.&~~~~~~~~&~&~&~&~&\\
~&~0&.&.&.&.&.&.&.\\
~&~~&0&1&2&3&4&5&6&d$\rightarrow$
\end{tabular}
\end{center}
\bigskip
\centerline{Table 3: The brane scan of superconformal groups admitting 
singletons.}
\bigskip
Note, however, that the $(d=3,D=11)$, $OSp(4|8)$ slot (written in 
boldface) occupies 
a privileged position in that the corresponding $D=11$ supergravity  
theory admits the $AdS_{4} \times S^{7}$ solution with $OSp(4|8)$ 
symmetry, whereas the other supergravities do not admit 
solutions with the superconformal group as a symmetry. For example, 
$D=10$ supergravity admits an $AdS_{3} \times S^{7}$ solution 
\cite{DTV,DGT}, but it does not have the full $[OSp(8|2)]^{2}¥$ symmetry 
because 
the dilaton is non-trivial and acts as a conformal Killing vector on 
the $AdS_{3}$. This is slightly mysterious, since the bulk theory has 
less symmetry than the boundary theory. We shall return to this in 
sections \ref{aristocrat} and \ref{Maldacena}.

\subsection{Many a time and oft}
\la{oft}

One might hope that a theory of everything should predict not only the
{\it dimensionality} of spacetime, but also its {\it signature}.  For 
example, 
quantum consistency of the superstring requires $10$ spacetime
dimensions, but not necessarily the usual $(9, 1)$ signature.  The
signature is not completely arbitrary, however, since spacetime
supersymmetry allows only $(9,1)$, $(5,5)$ or $(1,9)$.  Unfortunately,
superstrings have as yet no answer to the question of why our universe
appears to be four-dimensional, let alone why it appears to have
signature $(3,1)$. The authors of \cite{BD2,Classical,super} therefore 
considered a world with an
arbitrary number $T$ of time dimensions and an arbitrary number $S$ of 
space
dimensions to see how far classical supermembranes restrict not only 
$S + T$
but $S$ and $T$ separately.  To this end they also allowed an $(s,t)$ 
signature
for the worldvolume of the membrane where $s \leq S$ and $t \leq T$ but 
are
otherwise arbitrary.  The resulting allowed signatures are summarized 
on the ``brane-molecule'' of Table 4, where $R$, $C$, $H$ and $O$ denote 
real ($1+1$), complex ($2+2$), quaternion ($4+4$) and octonion 
($8+8$), respectively.

Moreover, it is not difficult to repeat the $AdS$ analysis of section
\ref{bootstrap} for arbitrary signatures, and to show that there is once 
again a one-to-one correspondence between supermembranes whose 
worldvolume theories are described by scalar supermultiplets and 
superconformal theories in Nahm's classification admitting singleton 
representations \cite{BD2,Classical}. 

\begin{center} 
\begin{tabular}{cccccccccccccc}
$S\uparrow$&&&&&&&&&&&~\\
11&.&&&&&&&&&&&~\\
10&.&O&&&&&&&&&&~\\
9&H&H/O&O&&&&&&&&&~\\
8&.&H&&&&&&&&&&~\\
7&.&H&&&&&&&&&&~\\
6&.&H&&&&O&&&&&~\\
5&C&C/H&H&H&H&H/O&O&&&&&~\\
4&.&C&&&&H&&&&&&~\\
3&.&R/C&&&&H&&&&&&~\\
2&.&R&~&&&H&~&~&~&O&&~\\
1&.&~&R&R/C&C&C/H&H&H&H&H/O&O&~\\
0&.&.&.&.&.&C&.&.&.&H&.&.~\\
~&0&1&2&3&4&5&6&7&8&9&10&11&$T\rightarrow$
\end{tabular}
\end{center}
\bigskip
\centerline{ Table 4: The brane-molecule}

At the time, we posed the obvious question of why the mathematics of 
supermembranes seems to allow these universes with more than one time 
dimension whereas the physical world seems to demand just one. 
This question has recently been answered by Hull \cite{Hull}, who 
claims that all these possible mathematical signatures are allowed 
physically and that, despite appearances, they are dual to one 
another.  Hull's resolution is both radical and conservative at the 
same time: it is radical in introducing universes with more than one 
time dimension into physics but conservative in saying that the only 
many-time universes we need worry about are those that are really 
one-time universes in disguise!
 
\subsection{Doubletons and tripletons revisited}
\la{revisit}

These early works focussed on scalar supermultiplets because these
were the only $p$-branes known in 1988 \cite{AETW}. However, with
the discovery in 1990 of Type $II$ $p$-brane solitons
\cite{CHS1,CHS2,HS,DLgauge,Luscan}, vector
and tensor multiplets were also seen to play a role. In particular, the
worldvolume fields of the self-dual Type IIB superthreebrane were shown to
be described by an $(n=4,d=4)$ gauge theory \cite{DLgauge}, which on the
boundary of $AdS_5$ is just the doubleton supermultiplet of the
superconformal group $SU(2,2|4)$! Thus one can after all entertain a 
$3$-brane-doubleton-supergravity bootstrap similar to the
membrane-singleton-supergravity bootstrap of section \ref{bootstrap}, 
and we may now draw the doubleton
brane scan of Table 5. Once again, the restriction to $d=4$ is 
consistent with renormalizability.  Note, however, that the $(d=4,D=10)$, 
$SU(2,2|4)$ 
slot
(written in boldface) occupies a privileged position in that the 
corresponding
$D=10$ Type $IIB$ supergravity admits the $AdS_5 \times S^5$ solution with 
$SU(2,2|4)$ symmetry, whereas the other supergravities do not admit 
solutions
with the superconformal group as a symmetry since, as discussed in section
\ref{aristocrat}, the dilaton is again non-trivial.

 \bigskip\bigskip
\begin{center}
\begin{tabular}{cccccccccc}
~&D$\uparrow$&&&&&&\\
~&11&.&~~~~~~&~~~~~~~~~~~~&&~~~~~~~~~~~~&~~~~~~~~~~~~&\\
~&10&.&~~~~~~&&&${\bf SU(2,2|4)}$&~&\\
~&9&.&~~~~~~&&&&&\\
~&8&.&~~~~~~&&&$SU(2,2|2)$&&\\
~&7&.&~~~~~~&&&&&\\
~&6&.&~~~~~~&&&&~&~\\
~&5&.&~~~~~~&&&&&\\
~&4&.&~~~~~~&&&$SU(2,2|1)$~&&\\
~&3&.&~~~~~~&&~&&&\\
~&2&.&~~~~~~&&&&&&\\
~&1&.&~~~~~~&~&~&~&~&\\
~&0&.&~~~~~~.&.&.&.&.&.\\
~&~&0&~~~~~~1&2&3&4&5&6&d$\rightarrow$
\end{tabular}
\label{doubletons}
\end{center}
\bigskip
\centerline{Table 5: The brane scan of superconformal groups 
admitting doubletons}
\bigskip\bigskip

Similarly, with the discovery of the $M$-theory fivebrane \cite{Gueven}, 
it was
realized \cite{GT} that the zero modes are described by an
$((n_+,n_-)=(2,0),d=6)$ multiplet with a chiral $2$-form, $8$ spinors and 
$5$
scalars, which on the boundary of $AdS_7$ is just the tripleton 
supermultiplet
of the superconformal group $OSp(6,2|4)$! (These zero modes are the 
same as
those of the Type $IIA$ fivebrane, found previously in \cite{CHS1,CHS2}). 
Thus one
can after all also entertain a $5$-brane-tripleton-supergravity bootstrap
similar to the membrane-singleton-supergravity bootstrap of section
\ref{bootstrap}. Thus we may now draw the tripleton
brane scan of Table 6. Note once again, however, that the $(d=6,D=11)$, 
$OSp(6,2|4)$ slot
(written in boldface) occupies a privileged position in that the 
corresponding
$D=11$ supergravity admits the $AdS_7 \times S^4$ solution with 
$OSp(6,2|4)$ symmetry, whereas the other supergravities do not admit 
solutions
with the superconformal group as a symmetry since, as discussed in section
\ref{aristocrat}, the dilaton is again non-trivial.
\begin{center}
\begin{tabular}{cccccccccc}
~&D$\uparrow$&&&&&&\\
~&11&.&~~~~~~&~~~~~~~~~~~~&~~~~~~~~~~~~&~~~~~~~~~~~~&~~~~~~~~~~~~&
${\bf OSp(6,2|4)}$\\
~&10&.&~~~~~~&&&&~&\\
~&9&.&~~~~~~&&&&&\\
~&8&.&~~~~~~&&&&&\\
~&7&.&~~~~~~&&&&&$OSp(6,2|2)$\\
~&6&.&~~~~~~&&&&~&~\\
~&5&.&~~~~~~&&&&&\\
~&4&.&~~~~~~&&&~&&\\
~&3&.&~~~~~~&&~&&&\\
~&2&.&~~~~~~&&&&&&\\
~&1&.&~~~~~~&~&~&~&~&\\
~&0&.&~~~~~~.&.&.&.&.&.\\
~&~&0&~~~~~~1&2&3&4&5&6&d$\rightarrow$
\end{tabular}
\label{tripletons}
\end{center}
\bigskip
\centerline{Table 6: The brane scan of superconformal groups 
admitting tripletons}
 
With the inclusion of branes with vector and tensor supermultiplets on 
their
worldvolume, another curiosity arises. Whereas the singleton brane
scan of Table 3 exhausts all the scalar branes and the tripleton brane scan 
of
Table 6 exhausts all the tensor branes, the doubleton brane scan of Table 5
is only a subset of all the vector branes \cite{super}. The Type 
$IIB$ 
$3$-brane is special because gauge theories are conformal only in $d=4$.  
So taking the brane-supergravity bootstrap idea seriously in 1988 would have
lead to the earlier discovery of the $M$-theory fivebrane and Type $IIB$ 
$3$-brane, but not the other Type $II$ branes.

\subsection{Near horizon geometry and p-brane aristocracy}
\la{aristocrat}

More recently, $AdS$ has emerged as the near-horizon geometry of black
$p$-brane solutions \cite{GT,DGT,GHT,DKL} in $D$ dimensions. The dual
brane, with worldvolume dimension ${\tilde d}=D -d-2$, interpolates
between $D$-dimensional Minkowski space $M_{D}¥$ and $AdS_{{\tilde d}+1}
\times
S^{d+1}$ (or $M_{{\tilde d}+1}\times S^{3}$ if $d=2$).  To see this, 
we recall that such branes arise generically as  
solitons of the following action \cite{lublack}:
\be
I=\frac{1}{2\kappa^{2}}¥ \int d^{D}¥x \sqrt{-g}\left[R-\frac{1}{2} 
(\partial 
\phi)^{2}¥-\frac{1}{2(d+1)!}e^{-\alpha \phi} F_{d+1}{}^{2}\right]
\ee
where $F_{d+1}¥$ is the field strength of a $d$-form potential 
$A_{d}¥ $ and $\alpha$ is the constant
\be
\alpha^{2}¥=4-\frac{2d \tilde d}{d+\tilde d}
\la{alpha}
\ee
Written in terms of the 
$(d-1)$-brane sigma-model metric $e^{-{\alpha/d} \phi}¥g_{MN}¥$, the 
solutions are \cite{lublack,DKL}
\[
ds^{2} =\Delta^{\frac{d-2}{d}} (-dt^{2} +d{\bf x}.d{\bf x})+
\Delta^{-2}dr^{2}¥+r^{2}¥d\Omega_{d+1}¥{}^{2}¥
\]
\[
e^{-2\phi }¥=\Delta^{\alpha}¥
\]
\be
F_{d+1}¥=db^{d}¥\epsilon_{d+1}¥
\la{branesolution}
\ee
where $d{\bf x}.d{\bf x}$ is the Euclidean $(\tilde d-1)$ metric, and
\be
\Delta=1-\left(\frac{b}{r}\right)^{d}¥
\ee
The near horizon geometry corresponds to $r \sim b$, and we make 
the change of variable 
\be
r=b\left(1+ \frac{\lambda}{d}\right)
\ee
in which case
\be
ds^{2}¥ =[\lambda^{\frac{d-2}{d}}(-dt^{2}¥+d{\bf x}.d{\bf x})
+\left(\frac{b}{d}\right)^{2}¥\lambda^{-2}¥d\lambda^{2}¥+b^{2}¥
d\Omega_{d+1}{}^2](1+O(\lambda))
\ee
Neglecting the $O(\lambda)$ terms, as before, and defining the new 
coordinate
\be
\lambda=e^{\frac{d}{b}\zeta}
\ee
we get
\[
ds^{2} \sim e^{\frac{d-2}{b}\zeta} (-dt^{2}¥+d{\bf x}.d{\bf x}) + 
d\zeta^{2}¥
+b^{2}¥d\Omega_{d+1}{}^{2}
\]
\[
\phi \sim -\frac{d\alpha}{2b}\zeta
\]
\be
F_{d+1}¥=db^{d}¥\epsilon_{d+1}¥
\la{nearhorizon}
\ee
Thus for $d\neq 2$ the near-horizon geometry is $AdS_{\tilde d +1}¥ \times 
S^{d+1}¥$. Note, however, that the gradient of the dilaton is 
generically non-zero and plays the role of a conformal Killing vector on 
$AdS_{\tilde d +1}$. Consequently, there is no enhancement of 
symmetry in the near-horizon limit. The unbroken supersymmetry remains 
one-half and the bosonic symmetry remains $P_{\tilde d}¥\times 
SO(d+2)$.
(If $d=2$, then (\ref{nearhorizon}) reduces to
\[
ds^{2}¥= (-dt^{2}¥+d{\bf x}.d{\bf x} + d\zeta^{2}¥)+b^{2}¥d\Omega_{3}{}^{2}  
\]
\[
\phi \sim -\frac{\alpha}{b}\zeta
\]
\be
F_{3}¥ \sim 2b^{2}¥\epsilon_{3}¥
\ee
which is $M_{\tilde d+1}¥\times S^{3}¥$, with a linear dilaton 
vacuum. The bosonic symmetry remains $P_{\tilde d}¥\times 
SO(4)$.)

Of particular interest are the ($\alpha=0$) subset of solitons for which 
the 
dilaton is zero or constant: the {\it non-dilatonic $p$-branes}. From 
(\ref{alpha}) we see that for single branes there are only 3 cases:
\[
D=11: d=6, \tilde d=3
\]
\[
D=10: d=4, \tilde d=4
\] 
\[
D=11: d=3, \tilde d=6
\]
which are precisely the three cases that occupied privileged positions on 
the
singleton, doubleton and tripleton branescans of Tables 3, 5 and 6. 
Then the near-horizon geometry coincides with the $AdS_{\tilde d +1}¥ 
\times 
S^{d+1}¥$ non-dilatonic maximally symmetric compactifications of the 
corresponding 
supergravities. The supersymmetry doubles and the bosonic symmetry is 
also enhanced to $SO(\tilde d,2) \times SO(d+2)$. Thus the total 
symmetry is given by the conformal supergroups $OSp(4|8)$, 
$SU(2,2|4)$ and $OSp(6,2|4)$, respectively. 

For bound states of $N$ singly charged branes, the constant $\alpha$ gets 
replaced by \cite{DR2,DR,KKLP}
\be
\alpha^{2}¥=\frac{4}{N}-\frac{2d \tilde d}{d+\tilde d}
\la{alpha2}
\ee
A non-dilatonic solution ($\alpha$=0) occurs for $N=2$: 
\[
D=6: d=2, \tilde d=2 
\]
which is just the dyonic string \cite{Rahmfeld}, of which the self-dual 
string 
\cite{lublack} is a special case, whose near-horizon geometry is 
$AdS_{3}¥\times S^{3}$. For $N=3$ we have
\[
D=5: d=2, \tilde d=1
\]
which is the 3-charge black hole \cite{T}, whose near-horizon geometry is 
$AdS_{2} \times S^{3}$, and
\[
D=5: d=1, \tilde d=2
\]
which is the 3-charge string \cite{T} whose near-horizon geometry is 
$AdS_{3} \times S^{2}$. For $N=4$ we have
\[
D=4: d=1, \tilde d=1
\]
which is the 4-charge black hole \cite{CT1,CT2}, of which the 
Reissner-Nordstrom  
solution is a special case \cite{DR2}, and whose near-horizon geometry 
is $AdS_{2} \times S^{2}¥$ \cite{FG}.

Thus we see that not all branes are created equal. A {\it $p$-brane 
aristocracy} obtains whose members are those branes whose near-horizon
geometries have as their symmetry the conformal supergroups. As an example
of a plebian brane we can consider the ten-dimensional  superstring:
\[
D=10: d=6, \tilde d=2
\]
whose near-horizon geometry is the $AdS_{3} \times S^{7}$ but with a 
non-trivial dilaton of section \ref{bootstrap} which does not have 
the conformal group $[OSp(8|2)]^{2}¥$ as its symmetry, even though 
this group appears in the $(D=10, \tilde d=2)$ slot on the singleton 
branescan of Table 3. In which case, of course, one may ask what role do 
these singletons play. We shall return to this in section 
\ref{Maldacena}.  

\subsection{The membrane at the end of the universe}
\la{universe}

As further evidence of the membrane/supergravity bootstrap idea, solutions 
of the
combined $D=11$ supergravity/supermembrane equations were sought for which 
the
spacetime is $AdS_4 \times M^7$ and for which the supermembrane occupies 
the
boundary of the $AdS_4$: {\it the Membrane at the End of the
Universe} \cite{BDPS,BDPS2,CKKTV}.

The bosonic sector of the supermembrane equation is
\be
\partial_{i}¥(\sqrt{-h}h^{ij}¥\partial_{j}¥X^{N}¥g_{MN}¥)+
\frac{1}{2}\sqrt{-h}h^{ij}¥\partial_{i}¥X^{N}¥\partial_{j}¥X^{P}¥
\partial_{M}¥
g_{NP}¥+
\frac{1}{3!}\epsilon^{ijk}¥\partial_{i}¥X^{N}¥\partial_{j}¥X^{P}¥
\partial_{k}¥X^{Q}¥
F_{MNPQ}¥=0
\la{brane}
\ee
where
\be
h_{ij}¥=\partial_{i}¥X^{M}¥\partial_{j}¥X^{N}¥g_{MN}¥
\ee
A membrane configuration will have residual supersymmetry if there exist 
Killing spinors $\epsilon(X)$ satisfying \cite{BDPS,BDPS2}
\be
\tilde D_{M}¥\epsilon=0,~~~~~~~~\Gamma\epsilon(X)=\epsilon(X)
\ee
where $\tilde D_{M}$ is the $D=11$ supergravity covariant derivative 
appearing in the gravitino transformation rule and $\Gamma$ is given 
by
\be
\Gamma=\frac{1}{3!}\sqrt{-h}\epsilon^{ijk}\partial_{i}¥X^{M}¥
\partial_{j}¥X^{N}¥
\partial_{k}¥X^{P}¥\Gamma_{MNP}
\ee

Let us denote the membrane 
worldvolume coordinates by $\xi^{i}=(\tau, \sigma,\rho)$. The original 
membrane 
at the end of the universe \cite{BDPS,BDPS2} was embedded in the 
$AdS_{4}¥$
geometry as 
\be
ds^{2}=¥-(1+a^{2}¥r^{2}¥)d\tau^{2}¥+(1+a^{2}¥r^{2}¥)^{-1}¥dr^{2}¥
+r^{2}¥(d\sigma^{2}¥+sin^{2}¥ \sigma d\rho^{2}¥)
\la{end}
\ee
and has topology $S^{1}¥ \times S^{2}¥$. Consequently, the $OSp(4|8)$ 
singleton action is the one given in (\ref{boundary}) with its scalar 
mass terms. Alternatively, one could take as the membrane at the end of 
the
universe to be the near-horizon membrane, which is embedded as
\be
ds^{2}=e^{4\zeta/b} (-d\tau^{2}¥+d\sigma^{2}+d\rho^{2}¥) + d\zeta^{2}¥
\la{near}
\ee
and has $M_{3}¥$ topology. It is still possible to associate an 
$OSp(4|8)$ action but this time it is defined over $M_{3}$ and has no 
scalar mass terms \cite{CKKTV,DFFFTT}. One can continue to call these 
fields 
``singletons'', of course, if by singleton one simply means anything 
transforming according to the $D(1/2,0)$ and $D(1,1/2)$ representations 
of
$SO(3,2)$.  A comparison of these two approaches is discussed in some 
detail in 
\cite{CKKTV}.

\subsection{Supermembranes with fewer supersymmetries. Skew-whiffing.}
\la{fewer}

So far we have focussed attention on compactifications to $AdS_{\tilde 
d+1}¥$ on 
round spheres $S^{d+1}$ which have maximal supersymmetry, but the 
supergravity 
equations admit infinitely many other compactifications on Einstein 
spaces $X^{d+1}$ which have fewer supersymmetries \cite{DNP}. Indeed 
generic 
$X^{d+1}$ have no supersymmetries at all\footnote{Thus in the early 
eighties, 
the most 
highly prized solutions were those with many supersymmetries. Nowadays, 
bragging rights seem to go those which have none!}. We note in this 
connection the {\it skew-whiffing theorem} \cite{DNP}, which states that 
for 
every $AdS_{\tilde 
d+1}¥$ compactification preserving supersymmetry, there exists one 
with no supersymmetry simply obtained by reversing the orientation of 
$X^{d+1}$ (or, equivalently, reversing the sign of $F_{d+1}$). The 
only exceptions are when $X^{d+1}$ are round spheres which preserve 
the maximum supersymmetry for either orientation. A 
corollary is that other {\it symmetric spaces}, which necessarily admit an 
orientation-reversing isometry, can have no superymmeties. Examples 
are provided by products of round spheres.
   
The question naturally arises as to whether these compactifications 
with fewer supersymmetries also arise as near-horizon geometries of 
$p$-brane solitons.  The answer is yes and the soliton 
solutions are easy to construct \cite{DLPS,ccdfft}. One simply makes 
the replacement
\be
d\Omega_{d+1}{}^{2}¥ \rightarrow d{\hat \Omega}_{d+1}{}^{2}¥
\ee
in (\ref{branesolution}), where $d{\hat \Omega}_{d+1}{}^{2}¥$ is the 
metric 
on an arbitary Einstein space $X^{d+1}$ with the same scalar curvature 
as the round $S^{d+1}$. The space need only be Einstein, it need not 
be homogeneous \cite{DLPS}. (There also exist brane solutions on Ricci 
flat 
$X^{d+1}$ \cite{DLPS} but we shall not discuss them here).  Note, 
however, that these non-round-spherical solutions do not tend to 
$(D-d)$-dimensional Minkowski space as $r\rightarrow \infty$. Instead 
the metric on the $(D-\tilde d)$-dimensional space transverse to the 
brane is asymptotic to a generalized cone
\be
ds_{D-\tilde d}¥{}^{2}¥=dr^{2}¥+r^{2}¥d{\hat \Omega}_{d+1}{}^{2}¥
\ee
and $(D-d)$-dimensional translational invariance is absent except when 
$X^{d+1}$ is a round sphere. The number of supersymmetries preserved 
by these $p$-branes is determined by the number of Killing spinors 
on $X^{d+1}$.

To illustrate these ideas let us focus on the eleven-dimensional 
supermembrane. The usual supermembrane interpolates between $M_{11}$   
and $AdS_{4}¥ \times$ round $S^{7}¥$, has symmetry $P_{3}¥ \times SO(8)$ 
and 
preserves $1/2$ of the spacetime supersymmetries for either orientation 
of the round $S^{7}¥$. Replacing the round $S^{7}¥$ by  generic 
Einstein spaces $X^{7}$ leads to membranes with symmetry $P_{3} 
\times G$, where G is the isometry group of $X^{7}$. For example 
$G=SO(5) \times SO(3)$ for the squashed $S^{7}$ \cite{ADP,DNP2}. For 
one orientation of $X^{7}$, they preserve $N/16$ spacetime 
supersymmetries where $1 \leq N \leq 8$ is the number of Killing 
spinors on $X^{7}$; for the opposite orientation they preserve no 
supersymmetries since then $X^{7}$ has no Killing spinors. For 
example, $N=1$ for the left-squashed $S^7$ owing to its $G_{2}¥$ 
holonomy \cite{ADP,DNP,DNP2}, whereas $N=0$ for the right-squashed 
$S^{7}$. However, all these solutions satisfy the same Bogomol'nyi
bound between the mass and charge as the usual supermembrane 
\cite{DLPS}.  Of course, skew-whiffing is not the only way to obtain vacua 
with less than maximal supersymmetry. A summary of known $X^{7}$, 
their supersymmetries and stability properties is given in \cite{DNP}. 
Note, however, that skew-whiffed vacua are automatically stable at the 
classical level since skew-whiffing affects only the spin $3/2$, $1/2$ 
and $0^{-}¥$ towers in the Kaluza-Klein spectrum, whereas the 
criterion for classical stability involves only the $0^{+}¥$ tower 
\cite{DNP}.

\subsection{$M$/$IIA$ duality and supersymmetry without supersymmetry in 
$AdS$}
\la{duality}

In more recent times, both perturbative and non-perturbative effects of 
ten-dimensional
superstring theory have been subsumed by an eleven-dimensional
theory \cite{Howe,Luduality,Hulltownsend,Townsendeleven,%
Wittenvarious,Duffliuminasian,Becker1,
Schwarzpower,Horava,TownsendM,Aharony,DuffM,BanksM}, called 
$M$-theory, whose low-energy limit is $D=11$ supergravity.  
In particular,
the $D=10$ Type $IIA$ superstring emerges from $M$-theory compactified
on $S^1$ \cite{Howe, Townsendeleven,Wittenvarious}.  In this picture,
the resulting Kaluza-Klein modes are Dirichlet $0$-branes
\cite{Polchinski} with masses proportional to $1/\lambda$ in the
string metric, where $\lambda$ is the string coupling constant.  They
are thus non-perturbative from the Type $IIA$ perspective.  This may
also be seen from the fact that perturbative string states carry no
Ramond-Ramond $U(1)$ charge whereas the massive Kaluza-Klein modes are
necessarily charged under this $U(1)$.  $M$-theory, on the other hand,
draws no distinction between perturbative and non-perturbative states.
An interesting question, therefore, is whether there is any difference
in the status of {\it supersymmetry} when viewed either from the
perturbative Type $IIA$ string or from the vantage point of
non-perturbative $M$-theory.  

This rehabilitation of $D=11$ supergravity has thus revived an 
interest in $AdS_{4}¥ \times X^{7}$ compactifications. In \cite{DLP}, 
for example, 
such $M$-theory vacua with $N>0$ supersymmetry were 
presented which, from the perspective of perturbative Type $IIA$ string 
theory, have $N=0$.
They can emerge whenever the $X^{7}$ is a
$U(1)$ bundle over a $6$-manifold.  The missing superpartners are
Dirichlet $0$-branes.  Someone unable to detect Ramond-Ramond charge
would thus conclude that these worlds have no unbroken supersymmetry.
In particular, the gravitinos (and also some of the gauge bosons) are
$0$-branes not seen in perturbation theory but which curiously remain
massless however weak the string coupling.  

The simplest example of this phenomenon is provided by the 
maximally-symmetric $S^7$ compactification 
\cite{DNP} of $D=11$ supergravity. Considered as a compactification of 
$D=11$ supergravity, the round
$S^7$ yields a four dimensional $AdS$ spacetime with $N=8$
supersymmetry and $SO(8)$ gauge symmetry, for either orientation of
$S^7$.  The Kaluza-Klein mass spectrum therefore falls into $SO(8)$
$N=8$ supermultiplets.  In particular, the massless sector is
described by gauged $N=8$ supergravity \cite{DNP}.  Since $S^7$ is a
$U(1)$ bundle over $CP^3$ the same field configuration is also a
solution of $D=10$ Type $IIA$ supergravity \cite{Nilssonpope}.  However, 
the
resulting vacuum has only $SU(4) \times U(1)$ symmetry and either $N=6$ or
$N=0$ supersymmetry depending on the orientation of the $S^7$.  The reason
for the discrepancy is that the modes charged under the $U(1)$ are
associated with the Kaluza-Klein reduction from $D=11$ to $D=10$ and
are hence absent from the Type $IIA$ spectrum originating from the
massless Type $IIA$ supergravity.  In other words, they are Dirichlet
$0$-branes and hence absent from the perturbative string
spectrum. There is thus more non-perturbative gauge symmetry and
supersymmetry than perturbative. (Here the words ``perturbative'' and
``non-perturbative'' are shorthand for ``with and without the inclusion
of Dirichlet $0$-branes'', but note that the Type $IIA$ compactification 
has
non-perturbative features even without the $0$-branes \cite{DLP}).  The
right-handed orientation is especially interesting because the perturbative
theory has no supersymmetry at all! See Table 7 (where we are using the
notation of \cite{slan} for $SU(4)$ representations). It is interesting  to
note that the $D=4$ massless states in the left-handed vacuum  originate
from the $n=0$ massless level and $n=1,2,$ massive Kaluza-Klein levels in 
$D=10$: whereas in the right-handed vacuum they originate from 
$n=0,1,2,3,4$ levels. 

\begin{center}
\begin{tabular}{|c|c|l|l|}\hline
Spin & $SO(8)$ reps & Left $SU(4) \times U(1)$ reps & Right $SU(4) 
\times U(1)$
reps\\  \hline\hline
$2$  & $1$    & $1_0$ & $1_0$ \\
$\ft32$& $8_s$  & $6_0+1_2+1_{-2}$ & $4_1+{\bar 4}_{-1}$ \\
$1$  & $28$   & $1_0+15_0+6_2+6_{-2}$ & $1_0+15_0+6_2+6_{-2}$ \\ 
$\ft12$& $56_s$ & $6_0+10_0+{\bar {10}}_0+15_2+15_{-2}$ & $4_1+{\bar
4}_{-1}+20_1+{\bar {20}}_{-1}+4_{-3}+{\bar 4}_{3}$ \\ $0^+$& $35_v$ &
$15_0+10_{-2}+{\bar {10}}_2$ & $15_0+10_{-2}+{\bar {10}}_2$ \\ $0^-$& 
$35_c$ &
$15_0+10_{2}+{\bar {10}}_{-2}$ & $1_0+{20'}_0+6_2+6_{-2}+1_4+1_{-4}$ \\ 
\hline
\end{tabular}
\end{center}
\centerline{Table 7: The massless multiplet under $SO(8) \rightarrow SU(4) 
\times U(1)$}
\label{massless}
\bigskip
A summary of perturbative versus 
non-perturbative symmetries is given in Table 8. In 
particular, the non-perturbative vacuum may have unbroken 
supersymmetry even when the perturbative vacuum has none.

\begin{center}
\begin{tabular}{|c|cc|cc|}\hline
Compactification& &Perturbative Type $IIA$& &Nonperturbative M-theory\\ 
\hline\hline
Left round $S^7$ & $N=6$ & $SU(4) \times U(1)$ & $N=8$ & $SO(8)$\\
Right round $S^7$ & $N=0$ & $SU(4) \times U(1)$ & $N=8$ & $SO(8)$\\
Left squashed $S^7$ & $N=1$ & $SO(5) \times U(1)$ & $N=1$ &  $SO(5) \times
SU(2)$\\ 
Right squashed $S^7$ & $N=0$ & $SO(5) \times U(1)$ & $N=0$ &  $SO(5) \times
SU(2)$\\
Left $M(3,2)$ & $N=0$ & $SU(3) \times SU(2) \times U(1)$ & $N=2$ & $SU(3) 
\times SU(2) \times U(1)$\\
Right $M(3,2)$ & $N=0$ & $SU(3) \times SU(2) \times U(1)$ & $N=0$ & $SU(3) 
\times SU(2) \times U(1)$\\ \hline
\end{tabular}
\end{center}
\bigskip
\centerline{Table 8: Perturbative versus non-perturbative symmetries}
\label{symmetries}
\bigskip

We cannot resist asking whether this could be a model of the real
world in which you can have your supersymmetry and eat it too
\footnote{A scheme in which you can have all the benefits of unbroken
supersymmetry while appearing to inhabit a non-supersymmetric world
has also been proposed by Witten \cite{Wittenworld} but his mechanism
is very different from ours. In particular, our vacua necessarily have
non-vanishing cosmological constant unless cancelled by fermion
condensates \cite{Orzalesi}.}. The problem with such a scenario, of
course, is that God does not do perturbation theory and presumably an
experimentalist would measure God's real world and not what a
perturbative string theorist thinks is the real world. Unless, for 
some unknown reason, the experimentalist's apparatus is so primitive as 
to be
unable to detect Ramond-Ramond charge in which case he or she would
conclude that the world has no unbroken supersymmetry.

\section{The new AdS/CFT correspondence}

\subsection{The Maldacena conjecture}
\la{Maldacena}

The year 1998 marks a revolution in anti de-Sitter space 
brought about by Maldacena's conjectured duality between physics in the 
bulk of $AdS$ and a conformal field theory on the boundary \cite{Maldacena}.
In particular, $M$-theory on $AdS_{4}¥\times S^{7}$ is dual to a 
non-abelian 
$(n=8,d=3)$ superconformal theory, Type $IIB$ string theory on 
$AdS_{5}¥\times S^{5}$ is dual to a $d=4$ $SU(N)$ super Yang-Mills theory 
theory and 
$M$-theory on $AdS_{7}¥\times S^{4}$ is dual to a non-abelian 
$((n_+,n_-)=(2,0),d=6)$ 
conformal theory. In particular, as has been spelled out most clearly in 
the $d=4$ 
$SU(N)$ Yang-Mills case, there is seen to be a correspondence between the
Kaluza-Klein mass spectrum in the bulk and the conformal dimension of
operators on the boundary \cite{Gubser,Witten}. 

One immediately recognises that the dimensions and supersymmetries of 
these 
three conformal theories are exactly the 
same as the singleton, doubleton and tripleton supermultiplets of 
section \ref{singletons}.  Moreover,  both the old and new $AdS/CFT$ 
correspondences are {\it holographic} in the sense of 
\cite{thooft,Susskind}. 
Following Maldacena's conjecture \cite{Maldacena}, therefore, 
a number of papers appeared reviving the old singleton-$AdS$-membrane-
superconformal field theory connections
\cite{Ferrara2,KKR,Boonstra,CKKTV,IMSY,Gunaydin,Gubser,%
Horowitz,Witten,Kachru,Berkooz,Ferrara,Rey,Maldacena2,lnv} and applying
them to this new duality context. What are the differences? 

One curious difference is that, with the exception of the three 
aristocratic branes, all the slots on the three brane-scans of superconformal 
field theories corresponded to bulk supergravities whose brane solutions  
are dilatonic, and hence have a symmetry smaller than the boundary 
theory. It seems that the branes at the end of the universe do not care
about the dilaton because the 
$r=$constant surfaces in (\ref{end}) (or the $\zeta=$ constant surfaces in 
(\ref {near})) posess the full superconformal symmetry even 
though the bulk $AdS$ solution does not. In other words, 
they admit the maximal set of conformal Killing vectors even though the  
the bulk admits less than the maximal set of Killing vectors. 
This contrasts 
with the new $AdS/CFT$ conjecture where a non-conformal supergravity 
solution 
in the bulk \cite{DGT} is deemed to be dual to non-conformal field theory 
on 
the boundary \cite{IMSY}. It is not obvious at the moment whether 
this difference is real or apparent and it would be interesting to 
pursue the matter further. 

Secondly, attention was focussed on {\it free} superconformal theories on 
the boundary as opposed to the {\it interacting} theories currently 
under consideration.  For example, although the worldvolume 
fields of the Type $IIB$ $3$-brane were known to be described by an 
$(n=4,d=4)$ gauge theory \cite{DLgauge}, we now know that this brane 
admits the interpretation of a Dirichlet brane \cite{Polchinski} and 
that the superposition of $N$ such branes yields a {\it non-abelian} 
$SU(N)$ gauge theory
\cite{witt}. These observations are crucial to the new duality conjecture 
\cite{Maldacena}. For earlier related work on coincident threebranes 
and $n=4$ super Yang Mills, see \cite{Gubser1,Klebanov2,Gubser2,Gubser3}.
Let us consider the solution for $N$ coincident $3$-branes 
corresponding to $N$ units of $5$-form flux \cite{HS,DLgauge}:
\[
ds^{2} =\Delta^{\frac{1}{2}} (-dt^{2} +d{\bf x}.d{\bf x})+
\Delta^{-2}dr^{2}¥+r^{2}¥d\Omega_{5}¥{}^{2}¥
\]
\be
F_{5}¥=4Nb^{4}¥\epsilon_{5}= *F_{5}
\la{3branesolution}
\ee
where $d{\bf x}.d{\bf x}$ is the Euclidean $3$-metric, and
\be
\Delta=1-\frac{Nb^{4}¥}{r^{4}¥}¥
\ee
Instead of regarding the near horizon geometry as an $r \sim 
N^{1/4}¥b$ limit we may equally well regard it as large $N$ limit, 
We find $AdS_{5}¥ \times 
S^{5}¥$, but with an $AdS$ radius proportional to $N^{1/4}$.
The philosophy is that Type $IIB$ supergravity is a good approximation for 
large $N$ and that Type $IIB$ stringy 
excitations correspond to operators whose dimensions diverge for 
$N \rightarrow \infty $. This makes contact with the whole industry of 
large $N$ QCD. These large $N$, non-abelian features were absent in the 
considerations of a $3$-brane/supergravity bootstrap discussed 
in section \ref{revisit}, as was the precise correspondence between the
Kaluza-Klein mass spectrum in the bulk and the conformal dimension of
operators on the boundary \cite{Gubser,Witten}. Nevertheless, as the 
present paper hopes to show, there are sufficently many similarities 
between the current bulk/boundary duality and the old Membrane at 
the End of the Universe idea, to merit further comparisons. 

 It is to the $AdS_{5}¥ \times S^{5}¥$ case that we now turn. Noting that
$T$-duality untwists  $S^{5}¥$ to $CP^2 \times S^1$, we construct the 
duality
chain $n=4$ super Yang-Mills $\rightarrow$ Type $IIB$ superstring on
$AdS_{5} \times S^{5}$ $\rightarrow$ Type $IIA$ superstring on $AdS_{5}
\times CP^{2}\times S^{1}$ $\rightarrow $ $M$-theory on AdS$_{5}
\times CP^{2}\times T^{2}$ \cite{DLP1}.  This provides another example of 
the 
phenomenon of {\it supersymmetry without
supersymmetry} \cite{DLP}, but this time without involving Dirichlet
$0$-branes.   On AdS$_{5} \times CP^{2} \times S^{1}$
Type IIA {\it supergravity} has $SU(3) \times U(1)\times U(1)\times
U(1)$ and $N=0$
supersymmetry. Indeed, since $CP^{2}$ does not admit a spin structure,
its spectrum contains no fermions at all! Nevertheless, Type IIA {\it
string theory} has $SO(6)$ and $N=8$ supersymmetry. The missing
superpartners (and indeed all the fermions) are provided by stringy
winding modes.  These winding modes also enhance $SU(3)\times U(1)$ to
$SO(6)$, while the gauge bosons of the remaining $U(1)\times U(1)$
belong to massive multiplets.

As a preliminary, we shall show how to construct the odd-dimensional 
unit spheres $S^{2n+1}¥$ as $U(1)$ bundles over $CP^{n}¥$

\subsection{Hopf fibrations}
\la{hopf}

The construction, which generalizes the $S^{7}$ example of section 
\ref{duality}, 
involves writing the metric $d\Omega_{2n+1}^2$ on the unit $(2n+1)$-sphere
in terms of the Fubini-Study metric $d\Sigma_{2n}^2$ on $CP^n$ as
\be
d\Omega_{2n+1}^2 = d\Sigma_{2n}^2 + (dz+\bar{\cal A})^2\ .\label{hopfscpn}
\ee
In fact we may give general results for any metric of the form
\be
ds^2 = c^2\, (dz + \bar{\cal A})^2 + d\bar s^2\label{bundle}
\ee
on a $U(1)$ bundle over a base manifold with metric $d\bar s^2$, where $c$
is a constant.  Choosing the vielbein basis $e^z = c \, (dz+\bar {\cal A}),
\, e^i = \bar e^i$, one finds that the Riemann tensor for $ds^2$ has
non-vanishing vielbein components given by
\bea
R_{ijk\ell } &=& \bar R_{ijk\ell} -\ft14 c^2(\bar {\cal F}_{ik}\, \bar{\cal
F}_{j\ell} - \bar{\cal F}_{i\ell}\, \bar{\cal F}_{jk}+ 2\bar{\cal F}_{ij}\,
\bar{\cal F}_{k\ell})\ ,
\nn\\
R_{zizj} &=&\ft14 c^2\, \bar{\cal F}_{ik}\, \bar{\cal F}_{jk}\ ,\qquad
R_{ijkz} = \ft12 c\, \bar\nabla_k\, \bar{\cal F}_{ij}\ .\label{curv}
\eea
In all the cases we shall consider, the components $R_{ijkz}$ will be zero,
since $\bar{\cal F}=d\bar{\cal A}$ will be proportional to
covariantly-constant tensors, such as K\"ahler forms.  The Ricci tensor for
$ds^2$ has the vielbein components
\be
R_{zz}=\ft14 c^2\, \bar{\cal F}_{ij}\, \bar{\cal F}_{ij}\ ,\qquad R_{ij}=
\bar R_{ij} 
-\ft12 c^2 \, \bar{\cal F}_{ik}\, \bar{\cal F}_{jk}\, \qquad 
R_{zi}= -\ft12 c\, 
\bar\nabla_j \bar{\cal F}_{ij}\ ,\label{kkscpn}
\ee

Applied to our present case, where the unit $(2n+1)$-sphere 
should have a Ricci tensor satisfying $R_{ab}=2n\,\delta_{ab}$, we see that 
this is achieved by taking the field strength to be given by 
$\bar{\cal F}_{ij} 
=2 J_{ij}$, where $J_{ij}$ is the covariantly-constant K\"ahler form on 
$CP^n$.  Furthermore, the Fubini-Study Einstein metric on $CP^n$ should be 
scaled such that its Ricci tensor satisfies $\bar R_{ij}=2(n+1)\, 
\delta_{ij}$.  The volume form $\Omega_{2n+1}$ on the unit $(2n+1)$-sphere
is related to the volume form $\Sigma_{2n}$ on $CP^{n}$ by
$\Omega_{2n+1}=dz\wedge \Sigma_{2n}$.  Note also that the volume form on 
$CP^n$ is related to the K\"ahler form by
\be
\Sigma_{2n} = \ft1{n!}\, J^n\ .
\ee

\subsection{$AdS_5\times S^5$ untwisted}
\la{twist}

Let us write the AdS$_5\times S^5$ geometry in the form
\bea
ds^2 &=& ds^2(AdS_5) + ds^2(S^5)\ ,\nn\\
H_\5 &=& 4m \Omega_{AdS_5} + 4m \Omega_{S^5}\ ,\label{2bconfig}
\eea
where $\Omega_{AdS_5}$ and $\Omega_{S^5}$ are the volume forms on
AdS$_5$ and $S^5$ respectively, $m$ is a constant, and the metrics on
AdS$_5$ and $S^5$ satisfy
\be
R_{\mu\nu} = -4 m^2\, g_{\mu\nu}\ ,\qquad 
R_{mn} = 4 m^2\, g_{mn}
\ee
respectively.  Since the unit 5-sphere has metric $d\Omega^2_5$
with Ricci tensor $\bar R_{mn} = 4\, \bar g_{mn}$, it follows that we 
can write 
\be
ds^2(S^5) = \ft1{m^2}\, d\Omega^2_5\ .
\ee
From (\ref{hopfscpn}), it follows that we can write this as 
\be
ds^2(S^5) = \ft1{m^2}\, d\Sigma_4^2 + \ft1{m^2}\, 
(dz+\bar{\cal A})^2\ ,
\ee
where $d\Sigma_4^2$ is the metric on the ``unit'' $CP^2$, and $d
\bar{\cal A}= 2J$, where $J$ is the K\"ahler form on $CP^2$.

     We may now perform a dimensional reduction of this solution to
$D=9$, by compactifying on the circle of the $U(1)$ fibres,
parameterized by $z$.  Comparing with the general Kaluza-Klein
prescription, for which
\bea
ds_{10}^2 &=& ds_9^2 + (dz_2 +{\cal A})^2\ ,\nn\\
H_\5 &=& H_\5 + H_\4\wedge (dz_2 + {\cal A})\ ,
\eea
we see, from the fact that the $S^5$ and $CP^2$ volume forms are
related by $\Omega_5 = (dz+\bar{\cal A})\wedge \Sigma_4$, that the
solution will take the 9-dimensional form
\bea
ds_9^2 &=& ds^2(AdS_5) + \ft1{m^2}\, d\Sigma_4^2\ ,\nn\\
F_\4 &=& \ft4{m^3}\, \Sigma_4\ ,\qquad {\cal F}_\2 = \ft2{m}\, J\ .
\label{d92bsol}
\eea
(Note that in the dimensional reduction of the 5-form of the type IIB
theory, its self-duality translates into the statement that the fields
$H_\5$ and $H_\4$ in $D=9$ must satisfy $H_\4 ={* H_\5} =F_\4$.)

     We now perform the $T$-duality transformation to the fields of the
$D=9$ reduction of the Type $IIA$ theory.  The relation between the 
$IIB$
and the $IIA$ fields is given in \cite{DLP1}.  Thus in the 
$IIA$
notation, we have the nine-dimensional configuration
\bea
ds_9^2 &=& ds^2(AdS_5) + \ft1{m^2}\, d\Sigma_4^2\ ,\nn\\
F_\4 &=& \ft4{m^3}\, \Sigma_4\ ,\qquad F_2^{(12)} = \ft2{m}\, J\ .
\label{2asol}
\eea
The crucial point is that the $2$-form field strength $F_2^{(12)}$ of
the IIA variables is no longer a Kaluza-Klein field coming from the
metric; rather, it comes from the dimensional reduction of the $3$-form
field strength in $D=10$.  Indeed, if we trace the solution
(\ref{2asol}) back to $D=10$, we have the Type $IIA$ configuration
\bea
ds_{10}^2 &=& ds^2(AdS_5) + \ft1{m^2}\, d\Sigma_4^2 + dz_2^2\ ,\nn\\
F_\4 &=& \ft4{m^3}\, \Sigma_4\ ,\qquad F_3^{(1)} = \ft2{m}\, J\wedge dz_2\ .
\label{2ad10}
\eea
The solution has the topology $AdS_5\times CP^2\times S^1$.  This
should be contrasted with the topology $AdS_5\times S^5$ for the
original $D=10$ solution in the Type $IIB$ framework.  Thus the
$T$-duality transformation in $D=9$ has ``unravelled'' the twisting of
the $U(1)$ fibre bundle over $CP^2$, leaving us with a direct product
$CP^2\times S^1$ compactifying manifold in the Type $IIA$ description.

 At first sight, the $T$-duality transformation that we have
performed has a somewhat surprising implication.  We began with a
solution on $AdS_5\times S^5$, which admits a spin structure, and
mapped it via $T$-duality to a solution on $AdS_5\times
CP^2\times S^1$, which does not admit a spin structure (because $CP^2$
does not admit a spin structure).  In particular, this means that the
spectrum of Kaluza-Klein excitations in the $CP^2\times S^1$
compactification of Type $IIA$ supergravity contains no fermions at all!
The equivalence is restored only when the stringy winding modes are
incorporated.  Further details may be found in \cite{DLP1}.

\subsection{Less supersymmetry}
\la{less}

Example of Type $IIB$ compactifications to $AdS_{5}¥$ with less 
supersymmetry, arising as in section \ref{fewer} from the near-horizon 
geometry of $3$-branes with less supersymmetry, may be obtained by   
replacing $S^{5}$  
by generic Einstein spaces $X^{5}$. Examples include: orbifolds of $S^{5}$ 
which 
can preserve $N=4,2,0$ \cite{Kachru,Oz}; non-singular lens spaces 
$S^{5}/Z_{n}$ 
which can preserve $N=4,2,0$ \cite{DLP,Halyo} (reducing the supersymmetry 
using lens 
spaces was discussed in \cite{DNP3}); $Q(n_{1},n_{2})$ spaces which are 
$U(1)$ 
bundles over $S^{2} \times S^{2}$ and which generically have $N=0$ but 
have $N=4$ for $(n_{1}¥ ,n_{2}¥)=(1,1)$ \cite{Page,DLP} (This 
leads to one of the gauged $(D=5,N=4)$ supergravities discussed in 
\cite{Romans2}); $T^{p,q}$ spaces which are cosets $[SU(2) \times 
SU(2)]/U(1)$ and which generically have $N=0$ but have $N=2$ for 
$(p,q)=(1,1)$ \cite{Romans,klebanov}.  

\subsection{$AdS_3\times S^3$ (un)twisted and squashed}
\la{squash}

As discussed in section \ref{aristocrat}, the six-dimensional space 
$AdS_{3}\times S^{3}$ emerges as the near
horizon geometry \cite{DGT,GHT} of the self-dual string
\cite{lublack,DKL} or, more generally, the dyonic string
\cite{Rahmfeld,DKL,DLPhet,DLPgauge}. The dyonic string admits the 
ten-dimensional
interpretation \cite{Rahmfeld} of an intersecting $NS-NS$ $1$-brane and
$5$-brane, which in a Type $II$ context is in turn related by $U$-duality to
the $D1-D5$ brane system
\cite{Boonstra1,Boonstra2,Horowitz2,Horowitz3,Maldacena3}. This
geometry plays a part in recent studies of black holes and has attracted a 
good deal of attention lately following Maldacena's conjecture.  $AdS_{3}$
is particularly interesting in this regard because the conformal field
theory on the boundary is then of the familiar and well-understood $1+1$ 
dimensional variety.

In this section, we wish to apply the above Hopf duality techniques to find 
Type
$IIA$ (and hence $M$-theory) duals of six-dimensional Type $IIB$ 
$AdS_{3}\times
S^{3}$ configurations obtained by either $T^{4}$ or $K3$
compactifications \cite{DLP2}. The novel ingredient is that these can be 
supported
by both $NS-NS$ and $R-R$ $3$-forms, in contrast to the $AdS_{5} \times
S^{5}$ example where the $5$-form was strictly $R-R$. This has some
interesting and unexpected consequences.  Noting that $S^{3}$ is a
$U(1)$ bundle over $CP^{1} \sim S^{2}$, we construct the dual Type 
$IIA$
configurations by a Hopf $T$-duality along the $U(1)$ fibre. In the case
where there are only $R-R$ charges, the $S^{3}$ is untwisted to
$S^{2}\times S^{1}$ (in analogy with the previous treatment of $AdS_{5}
\times S^{5}$).  However, in the case where there are only $NS-NS$
charges, the $S^{3}$ becomes the cyclic lens space $S^{3}/Z_{p}$ with
its round metric (and is hence invariant when $p=1$), where $p$ is the
magnetic $NS-NS$ charge.  In the generic case with $NS-NS$ and $R-R$
charges, the $S^{3}$ not only becomes $S^{3}/Z_{p}$ but is also
squashed, with a squashing parameter that is related to the values of
the charges.  Similar results apply if we regard $AdS_{3}$ as a bundle
over $AdS_{2}$ and $T$-dualize along the fibre.  We note that these Hopf
dualities preserve the area of the horizons, and hence they preserve
the black hole entropies.  

The dyonic string solution is supported
either by the $NS-NS$ $3$-form $F_\3^\ns$ or the $R-R$ $3$-form 
$F_\3^\rr$.
More general solutions can be obtained by acting with the $O(2,2)$
symmetry of the theory, allowing us, in particular, to find solutions
for dyonic strings carrying both $NS-NS$ and $R-R$ charges.  This in
done in detail in \cite{DLP2}, obtaining an $O(2,2;\Z)$ multiplet of dyonic
strings.

Near the horizon, even though the above dyonic solutions carry four
independent charges, the $3$-forms $F_\3^\ns$ and $F_\3^\rr$ become
self-dual, and the metric approaches that of $AdS_3\times S^3$.  The
dilatons $\phi_1$ and $\phi_2$ and the axions $\chi_1$ and $\chi_2$
are constant in the solution, and for simplicity we shall take them to
be zero.  The remaining equations are solved by taking the metric and
$3$-forms to be
\bea
ds_6^2 &=& ds^2(\ads) + ds^2(S^3)\ ,\nn\\
F_\3^\ns &=& \lambda\, \ep(\ads) + \lambda\, \ep(S^3)\ ,\label{adssol}\\
F_\3^\rr &=& \mu\, \ep(\ads) + \mu\, \ep(S^3)\ ,\nn
\eea
where $\lambda$ and $\mu$ are constants, and the metrics on the
AdS$_3$ and $S^3$ have Ricci tensors given by 
\be
R_{\mu\nu} = -\ft12(\lambda^2 +\mu^2)\, g_{\mu\nu}\ ,\qquad
R_{mn} = \ft12(\lambda^2 +\mu^2)\, g_{mn}\label{ricci}
\ee
respectively.  The
constants $\lambda$ and $\mu$ are related to the magnetic charges as
follows:
\be
Q_\ns \equiv \ft1{16\pi^2}
        \int F_\3^\ns = \fft{\lambda}{(\lambda^2 +\mu^2)^{3/2}}\ ,
\qquad
Q_\rr \equiv \ft1{16\pi^2}
     \int F_\3^\rr = \fft{\mu}{(\lambda^2 +\mu^2)^{3/2}}
\ .\label{charges}
\ee
We now make use of the fact that the metric $d\Omega_3^2$ can be
written as a $U(1)$ bundle over $CP^1\sim S^2$ as follows:
\be
d\Omega_3^2 = \ft14 d\Omega_2^2 +\ft14 (dz + B)^2 \ ,\label{fib32}
\ee
where $d\Omega_2^2$ is the metric on the unit 2-sphere, whose volume
form $\Omega_\2$ is given by $\Omega_\2= dB$.  (If $d\Omega_2^2$ is
written in spherical polar coordinates as $d\Omega_2^2 = d\theta^2 +
\sin^2\theta\, d\phi^2$, then we can write $B$ as $B=\cos\theta\,
d\phi$.) The fibre
coordinate $z$ has period $4\pi$.  Thus the six-dimensional metric
given in (\ref{adssol}) can be written as 
\be
ds_6^2 = ds^2(\ads) + \fft1{\lambda^2+\mu^2}\, d\Omega_2^2 +
\fft1{\lambda^2+\mu^2}\, (dz + B)^2 \ .\label{2bmet}
\ee
The four-dimensional area of the horizon is given by
\be A\sim L\, (\lambda^2+\mu^2)^{-3/2}\ ,\label{area} \ee
where $L$ is the contribution from $ds^2({\rm AdS})$ at the boundary at 
constant time.  The field strengths in (\ref{adssol}) can now be written as
\bea
F_\3^\ns &=& \lambda\, \ep(\ads) +
\fft{\lambda}{(\lambda^2+\mu^2)^{3/2}}\, \Omega_\2\wedge(dz +B)
\ ,\nn\\
F_\3^\rr &=& \mu\, \ep(\ads) +
\fft{\mu}{(\lambda^2+\mu^2)^{3/2}}\, \Omega_\2\wedge(dz + B)\ .
\label{fnfs}
\eea
If we dimensionally reduced on the fibre
coordinate we obtain the $5$-dimensional metric
\be
ds_5^2 = (\lambda^2+\mu^2)^{-1/3}\, ds^2(\ads) +
(\lambda^2+\mu^2)^{-4/3}\, d\Omega_2^2\ ,
\ee
while the new dilaton $\varphi$ is a constant, given by
\be
e^{\varphi/\sqrt6} = (\lambda^2+\mu^2)^{-1/3}\ .\label{varphi}
\ee
Comparing (\ref{fnfs}) with the reduction ans\"atze $F_\n \rightarrow
F_\n + F_{\sst{(n-1)}}\wedge (dz+B)$ for the field
strengths, we find that in $D=5$ we have
\bea
&&F_\3^\ns = \lambda\, \ep(\ads)\ ,\qquad
F_{\2 1}^\ns = \fft{\lambda}{(\lambda^2+\mu^2)^{3/2}}\, \Omega_\2
\ ,\nn\\
&&F_\3^\rr = \mu\, \ep(\ads)\ ,\qquad
F_{\2 1}^\rr = \fft{\mu}{(\lambda^2+\mu^2)^{3/2}}\, \Omega_\2
\ ,\label{f5}\\
&&{\cal F}_\2 = dB=\Omega_\2 \ .\nn
\eea

    We are now in a position to implement the $T$-duality transformation
from the Type $IIB$ description to the Type $IIA$ description in $D=5$.
using the dictionary of \cite{DLP2}. We find
\bea
&& F_\3= \lambda\, \ep(\ads)\ ,\qquad
  \cF_\2 = \fft{\lambda}{(\lambda^2+\mu^2)^{3/2}}\, \Omega_\2\ ,\nn\\
&&F_{\3 1} = -\mu\, \ep(\ads)\ ,\qquad
F_\2 = \fft{\mu}{(\lambda^2+\mu^2)^{3/2}}\, \Omega_\2\ ,\label{ft5}\\
&&F_{\21} = \Omega_\2\ .\nn
\eea
 From the duality dictionary \cite{DLP2} and (\ref{varphi}), together 
with the fact that
we are taking $\phi_1=\phi_2=0$ in the original Type $IIB$ solution, it
follows that the dilatons in the Type $IIA$ picture will be given by 
\be
e^{\varphi}= (\lambda^2+\mu^2)^{1/\sqrt6}\ ,\qquad
e^{\phi_1}=e^{\phi_2} = (\lambda^2+\mu^2)^{1/2}\ .
\ee

Finally, we can uplift the Type $IIA$ solution that we have just
obtained back to $D=6$, by retracing the standard Kaluza-Klein
reduction steps.  Doing so, we find that the six-dimensional metric in
the Type $IIA$ picture is
\be
ds_6^2 = (\lambda^2+\mu^2)^{-1/2}\, ds^2(\ads) +
(\lambda^2+\mu^2)^{-3/2}\, \Big[ d\Omega_2^2 +
\fft{\lambda^2}{\lambda^2 + \mu^2}\, (dz' + B)^2 \Big]\ ,\label{2amet}
\ee
where $B$ is a potential such that $\Omega_\2 = dB$, and the
coordinate $z'$ is related to $z$ by 
\be
z = \fft{\lambda}{(\lambda^2 +\mu^2)^{3/2}}\, 
z' = Q_\ns\, z'\ .\label{zrel}
\ee
It is
straightforward to verify that the area of the horizon of the metric
(\ref{2amet}) is the same as that before the Hopf $T$-duality
transformation, given by (\ref{area}).  The Type $IIA$ field strengths
in $D=6$ are given by
\bea
&&F_\4 = -\mu\, \ep(\ads)\wedge (dz+ \cA_\1) \ ,\qquad
F_\3 = \lambda\, \ep(\ads) + \Omega_\2\wedge (dz+\cA_\1)\ ,\nn\\
&&F_\2 = \fft{\mu}{(\lambda^2+\mu^2)^{3/2}}\, \Omega_\2\ ,
\eea
where 
\be
{\cal A}_\1 = \fft{\lambda}{(\lambda^2+\mu^2)^{3/2}}\, B = 
Q_\ns\, B\ .\label{bredef}
\ee

     We find that the charges carried by these field strengths are as
follows:
\bea
Q^\3_{\rm elec} &\equiv & \ft1{16\pi^2}\int_{S^3}\, e^{-\phi_1-\phi_2}\,
{*F_\3} = Q_\ns\ ,\nn\\
Q^\3_{\rm mag} &\equiv & \ft1{16\pi^2}\int_{S^3}\,
F_\3 = 1\ ,\nn\\
Q^\4_{\rm elec} &\equiv & \ft1{4\pi}\int_{S^2}\, e^{\fft12\phi_1-
\fft32\phi_2}\, {*F_\4} = - Q_\rr\ ,\nn\\
Q^\2_{\rm mag} &\equiv & \ft1{4\pi}\int_{S^2} \,
F_\2 = Q_\rr\ .\label{2acharges}
\eea
If the fibre coordinate $z'$ in (\ref{2amet}) had had the period
$4\pi$, then the topology of the compact 3-space would have been
$S^3$.  Since it is related to $z$ as given in (\ref{zrel}), and $z$
has period $4\pi$, it follows that $z'$ has period $4\pi/Q_\ns$, and
hence the topology of the compact 3-space is $S^3/Z_{Q_\ns}$, the
cyclic lens space of order $Q_\ns$.  On the other hand the magnetic charge
carried by the field strength $F_\3$ is equal to 1, having started, in
the original solution, as $Q_\ns$.  Furthermore, we can see 
from (\ref{2amet}) that the metric on the
lens space is not in general the ``round'' one, but is instead
squashed along the $U(1)$ fibre direction, with a squashing factor
$\nu$ given by
\be
\nu = \fft{\lambda}{\sqrt{\lambda^2 +\mu^2}} = 
\fft{Q_\ns}{\sqrt{Q_\ns^2 + Q_\rr^2}}\ .
\ee

We could have considered original solutions
in which the constant dilatons $\phi_1$ and $\phi_2$ were non-zero, in
which case the original electric and magnetic charges need not have
been equal.  The lens space after the Hopf T-duality transformation
will then be $S^3/Z_{Q_\ns^{\rm mag}}$.  Also, we can generalise the
starting point further by consider a solution on the product of
$AdS_3$ and the lens space $S^3/Z_n$, rather than simply AdS$_3\times
S^3$.  (From the lower-dimensional point of view, this corresponds to
giving the Kaluza-Klein vector a magnetic charge $n$ rather than 1.)
If we do this, then we find that a Type $IIB$ solution 
$AdS_3\times S^3/Z_n$ with 
charges $Q_\ns^{\rm elec}$, $Q_\ns^{\rm mag}$,
$Q_\rr^{\rm elec}$ and $Q_\rr^{\rm mag}$ will result, after the
$T$-duality transformation, in a Type $IIA$ solution $AdS_3\times S^3/Z_{Q_{\rm
    mag}^\ns}$ with charges
\be
Q^\3_{\rm elec}= Q_\ns^{\rm elec}\ ,\qquad
Q^\3_{\rm mag}= n\ ,\qquad
Q^\4_{\rm elec}= - Q_\rr^{\rm elec}\ ,\qquad
Q^\2_{\rm mag}= Q_\rr^{\rm mag}\ .\label{2acharges2}
\ee

Although the construction of conformal field theories with background
$R-R$ charges is problematical, there is an exact CFT duality
statement in the case of pure $NS-NS$ charge \cite{DLP2}. Namely, strings on
$S^3/Z_n$ with $3$-form flux $m$ are dual to strings on $S^3/Z_m$ 
with $3$-form flux $n$.

\subsection{BPS without BPS}

  The Type $IIA$ configuration (\ref{2ad10}) can be further uplifted 
  to $D=11$:
\bea
ds_{11}^2&=& ds^2(AdS_5) + \ft1{m^2}\, d\Sigma_4^2 + dz_1^2 +dz_2^2\ ,\nn\\
F_\4 &=& \ft4{m^3}\, \Sigma_4 - \ft2{m}\, J\wedge dz_1\wedge dz_2\ .
\label{2ad11}
\eea
The topology of this solution is $AdS_5\times CP^2\times T^2$.  This is 
just the near-horizon ($y\sim 0$) geometry of
the $M$-theory dual of the full Type $IIB$ $3$-brane \cite{DLP1}:
\[
ds_{11}¥{}^{2}¥=y^{2/3}¥\left[H^{-1/3}¥(dx^{\mu}¥dx_{\mu}+
r^{-2}¥(dz_{1}¥{}^{2}¥+dz_{2}¥{}^{2}¥))+H^{2/3}¥(dy^{2}¥+
d\Sigma_{4}¥{}^{2}¥)\right]
\]
\[
H=1+\frac{1}{4}Qy^{4}¥
\]
\be
F_{4}¥=2Qdz_{1}\wedge dz_{2} \wedge J+Q\Sigma_{4}¥
\la{BPS}
\ee
The interesting observation is that this provides a solution of 
$M$-theory which, according to the transformation rules of 
$D=11$ supergravity, preserves no supersymmetry. Yet we know in fact 
that it is BPS because it is just the Type $IIB$ $3$-brane in disguise. 
This is but one example of the more general phenomenon of {\it BPS 
without BPS} provided by Hopf duality. This reminds us (if we needed 
reminding) that there is more 
to $M$-theory than 
$D=11$ supergravity, and if we knew what the correct equations of 
$M$-theory were we should find that (\ref{BPS}) is indeed BPS.

One reason these $M$-theory duals of Type $IIB$ phenomena are interesting 
is that, in the AdS/CFT duality, Type $IIB$ 
supergravity with its Kaluza-Klein excitations is a good approximation for 
large $N$, and stringy excitations correspond to CFT operators whose 
dimensions diverge for as $N \rightarrow \infty$. But since the Hopf 
$T$ duality interchanges stringy and Kaluza Klein modes, the 
$M$-theory description may throw light on the finite $N$ regime.

Many theorists are understandably excited about the AdS/CFT 
correspondence because of what $M$-theory can teach us about 
non-perturbative QCD. In my opinion, however, this is, in a sense, a 
diversion from the really fundamental issue: What is $M$-theory? So my 
hope is that this will be a two-way process and that superconformal 
theories will also teach us more about $M$-theory.

\section{Acknowledgements}

It is a pleasure to dedicate this paper to Dick Arnowitt. I am grateful for 
conversations with  
Karim Benakli, Parid Hoxha, Hong Lu, Jianxin Lu, Rene Martinez Acosta,  
Chris Pope, Ergin Sezgin and Per Sundell.

\end{document}